\documentclass[floatfix,twocolumn,rmp,showpacs,superscriptaddress,a4paper,floatfix,aps]{revtex4}

\usepackage{dcolumn,graphicx,amsmath,amssymb,txfonts}

\begin{document}

\title{The networked seceder model:\\Group formation in social and
  economic systems}

\author{Andreas Gr\"{o}nlund}
\email{gronlund@tp.umu.se}
\author{Petter Holme}
\email{holme@tp.umu.se}
\affiliation{Department of Physics, Ume{\aa} University, 901 87
  Ume{\aa}, Sweden}

\begin{abstract}
  The seceder model illustrates how the desire to be different than
  the average can lead to formation of groups in a population. We turn
  the original, agent based, seceder model into a model of network
  evolution. We find that the structural characteristics our model
  closely matches empirical social networks. Statistics for the
  dynamics of group formation are also given. Extensions of
  the model to networks of companies are also discussed.
\end{abstract}


\maketitle

\section{Introduction}

Social networks have ``community structure''---actors (vertices) with
the same interests, profession, age (and so on), organize into tightly
connected subnetworks, or communities.~\cite{jazz,gir:alg,gui:sesi}
Subnetworks are connected into larger conglomerates into a
hierarchical structure of larger and more loosely connected
structures. Over the last few years the
issue of communities in social networks has ventured beyond sociology
into the area of physicists' network
studies~\cite{mejn:rev,ba:rev,doromen:rev}. The problem how to detect
and quantify community structure in networks has been the topic of a number
papers~\cite{mejn:commu,radi:comm,gir:alg}, whereas a few other have been
models of networks with
community structure~\cite{motter:sn,jin:mo,skyrms:mo}.
In these models, the common properties defining the community are
external to the network evolution (in the sense that an individual
does not choose the community to belong to by virtue of his or her
position in the network). In this paper we present a model where the
community structure emerge as an effect of the agents
personal rationales.
We do this by constructing a networked version of an agent based
model---the seceder model~\cite{sec1,sec2,sec3,sec4}---of social group
formation based on the assumption that people actively tries to be
different than the average. Independence and the desire to be
different plays an important role in social group
formation~\cite{kamp:at}, this might be even more important in the
social networking of adolescents. The important observation is that
few wants to be different than \textit{anyone} else, rather one
tries to affiliate to non-central group. This type of mechanisms are
probably rather ubiquitous, so the connotations of eccentricity are
not intended for the name of the model. (See Ref.~\cite{taro} for a
non-scientific account of the formation of youth sub-cultures by these
and similar premises.)

Another system where the networked seceder model can serve as a
model---or at least a direction for extension of present models (see
e.g.\ Ref.~\cite{mcp:dance}))---is networks where the vertices are
companies and the edges indicate a similar niche. (Such edges can be
defined indirectly using stock-price correlations~\cite{bonanno:eco}.)
The establishment of new companies are naturally more frequent in new
markets. Assuming new markets are remote to more traditional markets,
the networked seceder model makes a good model of a such company
networks.

\section{Preliminaries}

\subsection{Notations}

The model we present produces a sequence of graphs $\{G_t\}$. Each
member of this sequence consists of the same set $V$ of $N$ vertices,
and a time specific set of $M$ undirected edges $E_t$. The model
defines a Markov process and is thus suitable for a Monte Carlo
simulation. The number of iterations of the algorithm defines the
simulation time $t=1,\cdots, t_\mathrm{max}$.

We let $d(i,j)$ denote the distance (number of edges in the shortest
path) between two vertices $i$ and $j$. We will also need the
\textit{eccentricity} defined as the maximal distance from $i$
to any other vertex.

\begin{figure}
  \centering{\resizebox*{0.79 \linewidth}{!}{\includegraphics{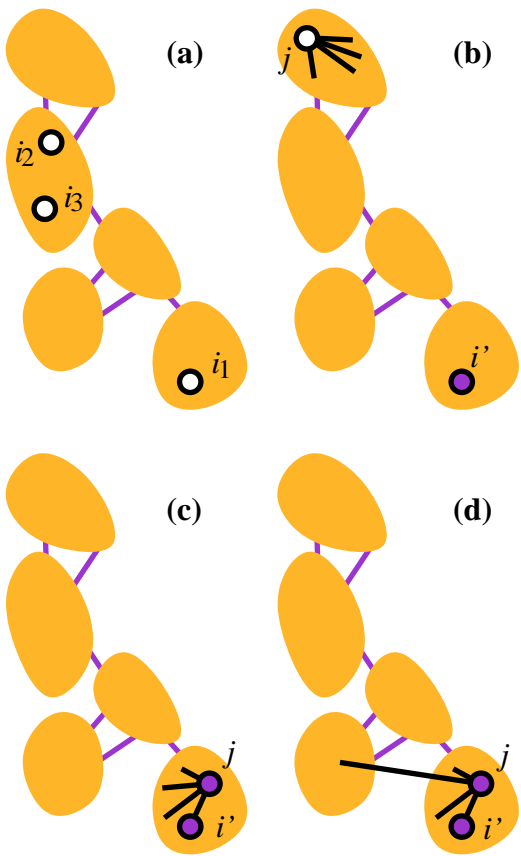}}}
  \caption{Illustration of the networked seceder model. (a) In
  step~\ref{lab:select} three vertices, $i_1$, $i_2$ and $i_3$, are
  chosen at random. (b) In step~\ref{lab:pick} the least central of
  the three vertices is relabeled to $\hat{i}$. In step \ref{lab:rewire}
  a vertex $j$ is selected at random and (c) the edges of $j$ are
  rewired to $\hat{i}$ and $\hat{i}$s neighborhood (and to a set of
  random other vertices if necessary. Note that, in (c), $j$ is moved to the
  cluster it is rewired to. In step~\ref{lab:random} $j$'s edges are
  rewired with a probability $p$. The shaded areas represent tightly
  connected subgraphs.}
  \label{fig:mo}
\end{figure}

\begin{figure*}
  \centering{\resizebox*{0.85 \linewidth}{!}{\includegraphics{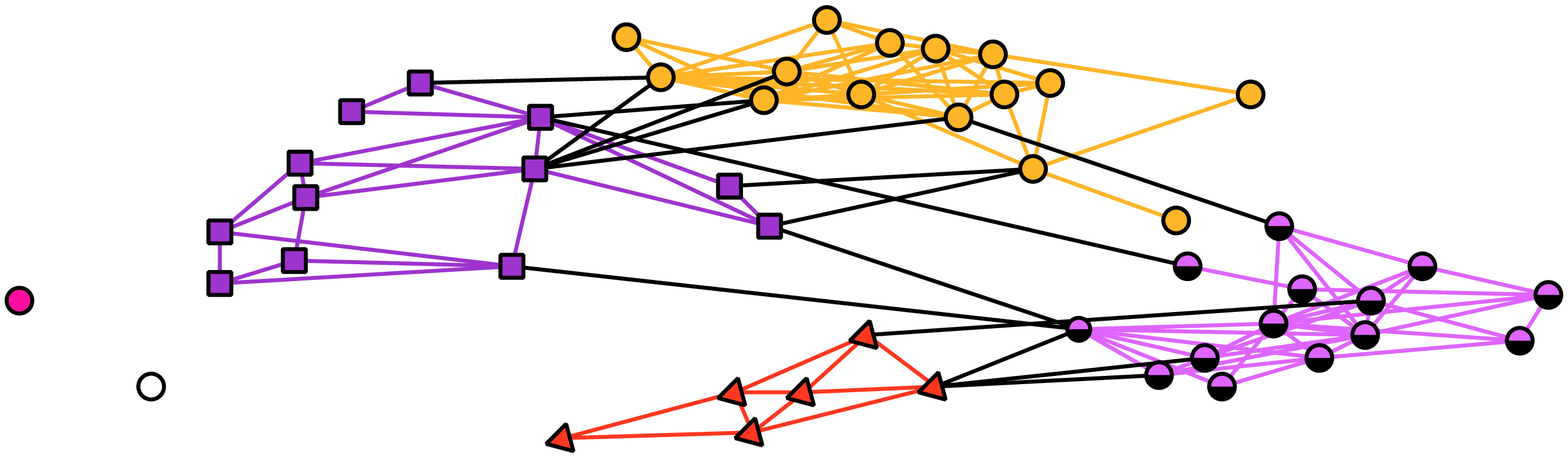}}}
  \caption{One realization of the networked seceder model. The model
    parameters are $N=50$, $M=150$ and $p=0$. The indicated groups are
    identified with Newman's clustering algorithm (see
    Sect.~\ref{sec:clust}). This realization have modularity
    $Q=0.575$, clustering coefficient $C=0.530$, and assortative mixing
    coefficient $r=0.0456$.}
  \label{fig:shot}
\end{figure*}

\subsection{The seceder model}

The original seceder model~\cite{sec1} is based on $N$ individuals
with a real number $s(i)$ representing the traits (or personality) of
individual $i$. The algorithm is then to repeat the following steps:
\begin{enumerate}
\item Select three individuals $i_1$, $i_2$ and $i_3$ with uniform
  randomness.
\item Pick the one (we call it $\hat{i}$) of these whose $s$-value is
  farthest away from the average $[s(i_1) + s(i_2) + s(i_3)]/3$.
\item Replace the $s$-value of a uniformly randomly chosen agent with
  $s(\hat{i})+\eta$, where $\eta$ is a random number from the normal
  distribution with mean zero and variance one. 
\end{enumerate}
Note that the actual values of $s$ is irrelevant, only the differences
between $s$ of different agents. The output of the seceder model is a
complex pattern of individuals that stick together in well-defined
groups. The groups has a life-cycle of their own---they are born,
spawn new groups and die. Statistical properties of the model is
investigated in Ref.~\cite{sec1}, effects of a bounded trait-space is
studied in Ref.~\cite{sec2}, the fitness landscape is the issue of
Ref.~\cite{sec3}, and Ref.~\cite{sec4} presents a generalization to
higher-dimensional trait-spaces.

Our generalization of this model to a network model based on the idea
that if the system is embedded in a network, then the difference in
personality is
implicitly expressed through the network position, so the identity
number (or vector) $s$ becomes superfluous. I.e., the homophily
assumption~\cite{mcp:bird}---that like attracts like---means that the difference in
character between two vertices $i$ and $j$ (defined as $|s(i)-s(j)|$
in the traditional seceder model) can be estimated by the graph
distance $d(i,j)$ in a networked model. The model we propose is then,
starting from any graph with $N$ vertices and $M$ edges, to iterate
the following steps:
\begin{enumerate}
\item Select three different vertices $i_1$, $i_2$ and $i_3$ with
  uniform randomness. \label{lab:select}
\item Pick the one $\hat{i}$ of these that is least central in the
  following sense: If the graph is connected vertices of highest
  eccentricity are the least central. If the graph is disconnected the
  most eccentric vertices within the smallest connected subgraph
  are the least central. If more than one vertex is least central, let
  $\hat{i}$ be a vertex in the set of least central vertices chosen
  uniformly randomly. \label{lab:pick}
\item Choose a vertex $j$ by uniform randomness. If $\deg j \leq \deg
  \hat{i} + 1$, rewire $j$'s edges to $\hat{i}$ and a random selection
  of $\hat{i}$'s  neighbors. If $\deg j \geq \deg \hat{i} + 1$, rewire $j$'s
  edges to $\hat{i}$,  $i$'s neighborhood and (if $\deg j > \deg
  \hat{i} + 1$) to $\deg j - \deg \hat{i} - 1$ randomly selected
  other vertices. \label{lab:rewire} \item Go through $j$'s edges once
  more and rewire these with a probability $p$ to a randomly chosen
  vertex. \label{lab:random}
\end{enumerate}
The rewiring of steps \ref{lab:rewire} and \ref{lab:random} are
performed with the restriction that no multiple edges or loops (edges
that goes from a vertex to itself) are allowed. Steps
\ref{lab:select} to \ref{lab:rewire} correspond rather closely to
the same steps of the original model. That $j$'s edges are rewired
mainly to the neighborhood of $\hat{i}$ (and $\hat{i}$ itself) reflects the
inheritance of trait value of the original model---by the homophily
assumption the neighborhood of $\hat{i}$ will have much the same
personality as $\hat{i}$ itself. The main difference between original and
the networked seceder model is step \ref{lab:random} where some
vertices are rewired to distant vertices. The motivation for this step
is that long-range connections exists in real-world
networks~\cite{wattsstrogatz,watts:small2}, and can in some situations
be even more important than the strong links of a cohesive
group~\cite{grano:weak}. This kind of rewiring, to obtain long-range
connections has been used to model ``small-world behavior'' of
networks~\cite{wattsstrogatz} (i.e.\ a logarithmic, or slower, scaling
of the average inter-vertex distance for ensembles of graphs with the
same average degree~\cite{mejn:rev}).

To make the model consistent we also have to specify the initial
graph. As far as we can see, at least for finite $p$, this choice is
irrelevant---the structure of the generated graphs are the same (or at
least very similar). We will not investigate this point
further. Instead we fix the initial graph to an instant of Erd\"{o}s
and R\'{e}nyi's random graph model~\cite{er:on} (for a modern survey
of this model, see Ref.~\cite{janson}): A graph with $N$ edges and $M$
edges is constructed by starting from isolated vertices and then
iteratively introduce edges between vertex-pairs chosen by uniform
randomness and with the restriction that no multiple edges or loops
are allowed. To be sure that the structure of the random graph is gone
we run the construction algorithm $10 N$ sweeps through every vertex
before the graph is sampled. (We justify this number \textit{a
  posteriori} below.)

An illustration of the construction algorithm can be seen in
Fig.~\ref{fig:mo}. An realization of the algorithm is displayed in
Fig.~\ref{fig:shot}. The $p$-value of this realization is zero. For
the value $p=0.1$ we use in most simulations the community structure
is less visible to the eye. Nevertheless---as we will see---the
community structure is still substantial for much larger values of $p$.

\subsection{Detecting communities\label{sec:clust}}

To analyze the structure of cohesive subgroups in our model networks
we use the community detection scheme presented in
Ref.~\cite{mejn:fast}. This algorithm starts from one-vertex clusters
and (somewhat reminiscent of the algorithm in Ref.~\cite{bohu:soc}) iteratively
merges clusters to form clusters of increasing size with relatively
few edges to the outside. The crucial ingredient in the scheme is a
quality function
\begin{equation}
  Q'=\sum_{s\in S} (e_{ss}-a_s^2)
\end{equation}
where $S$ is the set of subnetworks at a specific iteration of the
algorithm and $e_{ss'}$ is the fraction of edges that goes between a 
vertex in $s$ and a vertex in $s'$, and $a_s=\sum_{s'}e_{ss'}$. The
algorithm performs a steepest-accent in $Q'$-space---at each iteration
the two clusters that leads to the largest increase (or smallest
decrease) in $Q'$ are merged. The iteration having the highest $Q'$
value---which defines the \textit{modularity} $Q$---gives the
partition into subgroups.

\subsection{Conditional uniform graph tests}

One can argue that some network structures are more basic than
other. Given such an assumption and a network $G$, an interesting
issue is whether a certain structure, say $X$, is an artifact of a more
basic structure, say $Y$. One way to do this is by a
\textit{conditional uniform graph test}: One compares the value of
$X(G)$ with $X$ averaged over an ensemble of graphs with a the value of
$Y$ fixed to $Y(G)$. This has (since Ref.~\cite{katz:cug}) been a well
established technique in social network analysis and has recently been
brought over to physicists'~\cite{maslov:inet} and
biologists'~\cite{alon} network literature. A common
assumption~\cite{alon,maslov:inet,roberts:mcmc} is that the degree
distribution is such a very basic structure. We make this assumption
too and perform a conditional uniform graph test with respect to the
degree sequence of the networks. To sample networks with a given
degree sequence we use the idea of Ref.~\cite{roberts:mcmc} to rewire
the edges of the network in such a way that the degree sequence
remains unaltered. More precisely we go through all edges $(i,j)\in E$
and perform the following:
\begin{enumerate}
\item Construct the set $E'$ of edges such that if $(\hat{i},\hat{j})\in E'$
  then replacing $(i,j)$ and $(\hat{i},\hat{j})$ by $(i,\hat{j})$ and $(\hat{i},j)$ would
  not introduce any loops (self-edges) or multiple edges.
\item Pick an edge $(\hat{i},\hat{j})\in E'$ by uniform randomness.
\item Rewire $(i,j)$ to $(i,\hat{j})$ and $(\hat{i},\hat{j})$ to $(\hat{i},j)$.
\end{enumerate}

For every realization of the seceder algorithm we sample
$n_\mathrm{sample}=10$ randomized reference networks as described
above. The motivation for this
rather low number is that all quantities seems to be self-averaging
(the fluctuations decrease with $N$) and many have symmetric
distributions with respect to rewirings (which makes many realization
averages compensate for few rewiring averages). To further motivate
this small $n_\mathrm{sample}$ we compare with $n_\mathrm{sample}=100$
for the smallest size ($N=200$, which, as mentioned, is most affected
by fluctuations) and find that the quantities typically differ by
$0.5\%$ which we consider small.

\section{The community structure of the seceder model\label{sec:commu}}

\begin{figure}
  \centering{\resizebox*{\linewidth}{!}{\includegraphics{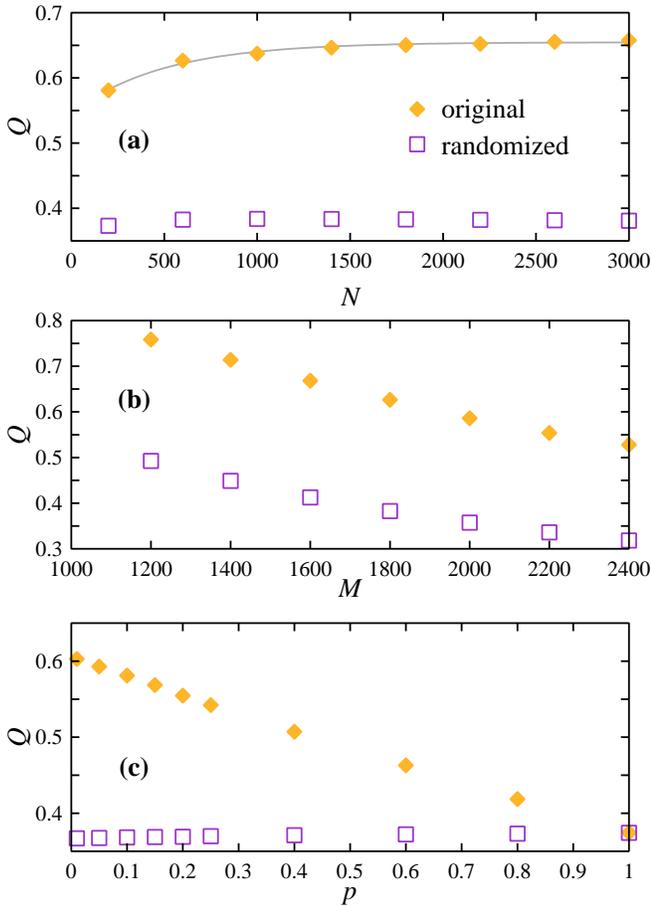}}}
  \caption{The modularity $Q$ as a function of the model
  parameters. (a) shows $Q$ as a function of $N$ with $M=3N$ and
  $p=0.1$. (b) displays $Q$ for different $M$ for $N=600$ and
  $p=0.1$. In (c) we plot the $p$ dependence of $Q$ for $N=200$ and
  $M=600$. The gray line in (a) is a fit to a exponential. All
  errorbars are smaller than the symbol size.}
  \label{fig:q}
\end{figure}

\begin{figure}
  \centering{\resizebox*{\linewidth}{!}{\includegraphics{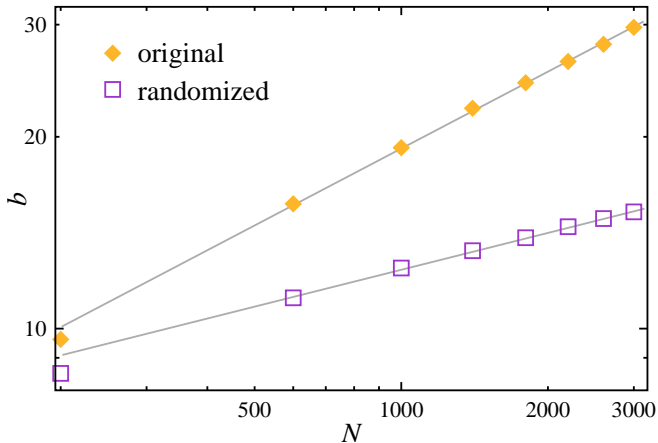}}}
  \caption{The number of groups $b$ as a function of the systems
    size $N$. The other parameter values are $M=3N$ and $p=0.1$. The line is
    a fit to a power-law $a b^\beta$. For this set of parameters
    $\beta=0.400(6)$ for the seceder model and $0.193(6)$ for the
    reference-graphs of the conditional uniform graph test. All
    errorbars are smaller than the symbol size. Note the
    double-logarithmic scale.}
  \label{fig:groups}
\end{figure}

The key quantity capturing the degree of community order in the
network is the modularity $Q$ (defined in Sect.~\ref{sec:clust})). In
Fig.~\ref{fig:q}(a) we see that, if the average degree and $p$ is kept
constant $Q$ converges to a high value, $Q\approx 0.64$ for $p=0.1$
and $M=3N$. This value is much higher than the reference value from
the randomized networks---this curve has a peak around $N=1500$ and
decays for larger $N$, larger sizes would be needed to see of $Q$
converges to a finite value for the randomized networks. With the analogy
to the Watts-Strogatz model (where
a fraction $p$ of a circulant's~\cite{harary} edges is rewired
randomly) we would say that $p=0.1$ is a rather high value, still $Q$
is much higher for the networked seceder model than for random
networks with the same degree distribution. From this we conclude that
our model fulfills its purpose---it produces networks with a pronounced 
community structure just as the original seceder model makes agents
divide into well-defined groups in trait-space. In Fig.~\ref{fig:q}(b)
we plot the $M$-dependence of $Q$ for fixed $N=600$ and $p=0.1$. We
see that $Q$ decreases with $M$ for both the seceder model and the
randomized networks. As $M$ approaches its maximum value $N(N-1)/2$ the
curves will converge (since the fully connected graph is unique), but
the figure shows that the curves are separated for a wide parameter
range. More importantly it suggests that the quantity $Q$ should be
rescaled by some appropriate function if networks of different average
degree are to be compared. In the rest of our paper, however, we will
keep the degree constant. In Fig.~\ref{fig:q}(c) we show the
$p$-dependence of $Q$. As expected $Q$ decays monotonously, in fact
almost linearly, with $p$. The curves for the seceder model converges
to the curve of the randomized networks as $p\rightarrow 0$. $Q$ of the
randomized reference networks is almost $p$-independent. The fact that
it is not completely $p$-independent means that the degree
distribution of the seceder model must vary with $p$. We will
strengthen this claim later.

Fig.~\ref{fig:groups} shows the size-dependence of $b$---the number of
groups. We see that this function can be well-described by a constant
plus a power-law,
\begin{equation}
A+b^\beta~,
\end{equation}
(where $A$ is a constant) with an exponent $\beta=0.400(6)$ for the
seceder model and $\beta=0.193(6)$ for
the random networks with the same degree distribution. The average
community-size is given by $N/b$ and will therefore also behave as a
power-law, with exponent $1-\beta=0.600(6)$. This fact that for the
number and average size of the communities grows with $N$ does not
seem contradictory to the real world to us. Since a community, both in
a social and economical interpretation of the model, does not need to
be controlled or supervised there is no natural upper limit to the
number of community members. Furthermore, there is no particular
constraint on the number of communities present in real world
systems. A thorough study of the scaling-exponents would be
interesting, but falls out of the scope of the present paper.

\begin{figure*}
  \centering{\resizebox*{0.7 \linewidth}{!}{\includegraphics{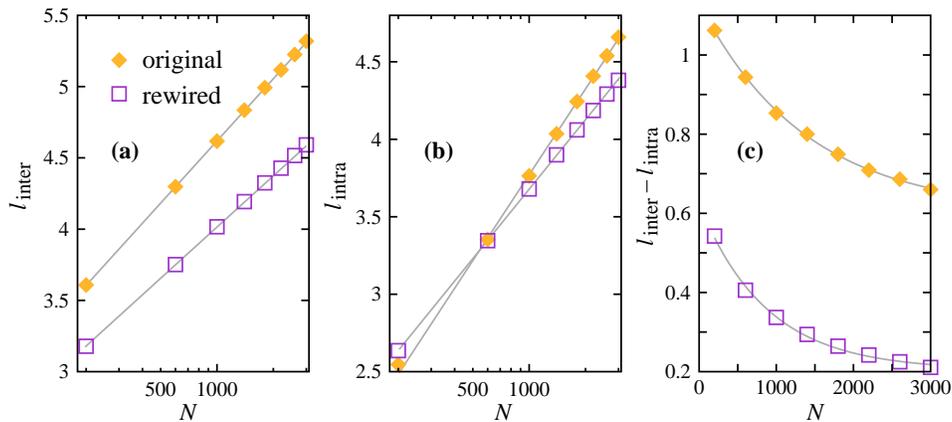}}}
  \caption{Average distance between and within clusters (as identified
    by the algorithm described in Sect.~\ref{sec:clust}). The gray
    lines are fits to an exponential form. The slope of the original
    is the same in (a) and (b) (also the rewired line has the same
    slope in (a) and (b)). All errorbars are smaller than the symbol
    size.}
  \label{fig:ii}
\end{figure*}

In Fig.~\ref{fig:ii} we display the average geodesic lengths within
a community $l_\mathrm{intra}$ and between vertices of different communities
$l_\mathrm{inter}$ for parameter values $M=3N$ and $p=0.1$. To be
precise, we consider the largest connected component (which typically
contains $99\%$ of the vertices), and define
\begin{subequations}
\begin{eqnarray}
  l_\mathrm{intra}=\frac{1}{N_\mathrm{intra}}\sum_{i=1}^b\sum_{v,w\in
  B^i} d(v,w) \mbox{~and}\\
  l_\mathrm{inter}=\frac{1}{\dbinom{N}{2}-N_\mathrm{intra}}
  \sum_{i=1}^b\sum_{v\in
  B^i}\sum_{w\notin  B^i} d(v,w)
\end{eqnarray}
\end{subequations}
where $B^i$ is the $i$'th cluster and
\begin{equation}
N_\mathrm{intra} = \sum_{i=1}^b \dbinom{i}{2}
\end{equation}
is the number of pairs of vertices belonging to the same community. As
seen in  Fig.~\ref{fig:ii}(a) and (b) both $l_\mathrm{intra}$ and
$l_\mathrm{inter}$ grows logarithmically as functions of $N$ with the
same slope in a semi-logarithmic plot. A logarithmic scaling of the
average shortest path length (which of course also holds) is expected (cf.\
Ref.~\cite{bollobas_chung_88}). But we could not 
anticipate the lack of qualitative difference between distances
between vertices of the same an different clusters. The actual values
of $l_\mathrm{intra}$ is significantly smaller than $l_\mathrm{inter}$
and this difference holds as $N\rightarrow\infty$: As seen in
Fig.~\ref{fig:ii}(c) $l_\mathrm{inter}-l_\mathrm{intra}$ converges to
$0.60(1)$. The same  value for the randomized graphs is
$l_\mathrm{inter}-l_\mathrm{intra}=0.204(8)$ which is expected---the
detected communities in the networked seceder model are more
well-defined and tight-knit that the corresponding communities in a
random network with the same degree distribution.

\section{Other structural characteristics\label{sec:osc}}

Apart from the quantities of the previous section, all directly
related to the community structure, we also look at some other well
established structural measures: The clustering coefficient, the
assortative mixing coefficient and the degree distribution.

\begin{figure}
  \centering{\resizebox*{0.9
  \linewidth}{!}{\includegraphics{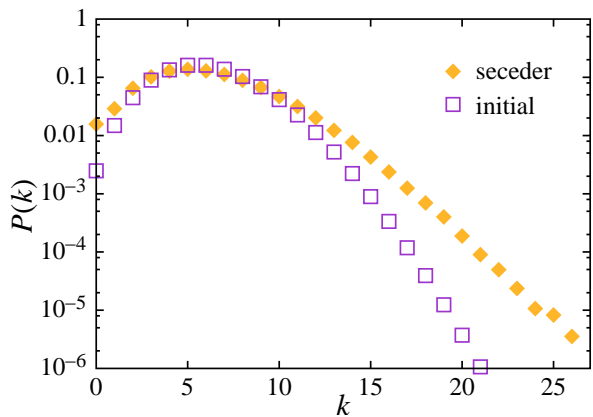}}} 
  \caption{Degree distribution of the networked seceder model. The
  model parameters are $N=1800$, $M=5400$ and $p=0.1$. The squares
  indicate the degree distribution of a random graph with the sizes
  ($N$ and $M$), i.e., the initial network before the iterations of
  the seceder model commence---see Eq.~(\ref{eq:poi}).} 
  \label{fig:deg}
\end{figure}

\subsection{Degree distribution}

Following the works of Barab\'{a}si and
coworkers~\cite{ba:model,alb:attack,bara:www} the degree distribution
has been perhaps the most studied network structure. Many of these
studies have found a skewed, power-law tailed,
degree-distribution. In some social networks---of telephone
calls~\cite{aiello}, e-mail communication~\cite{bornholdt:email}
and the network of sexual contacts~\cite{liljeros:sex}---authors have
found large tails of the degree distribution that fits well to
a power-law functional form. Other social network studies report degree
distributions with large degree cut-offs, these contain network of movie
actor~\cite{amaral:classes}, scientific
collaborations~\cite{mejn:scicolpnas} or Internet community
interaction~\cite{pok} or romantic interaction among High School
students (the network of Ref.~\cite{addh} as studied in
Ref.~\cite{mejn:rev}). Yet other studies have found social networks
with Gaussian degree distributions (the
acquaintance networks of Refs.~\cite{fararo} and \cite{ber:mormon}
studied in Ref.~\cite{amaral:classes}), or exponential degree
distributions (of e-mail networks~\cite{mejn:email,gui:sesi}). We
conclude that the degree distribution of social networks is still an
open question with, most likely, not a single solution---different
social networks may follow different degree distributions. The degree
distribution of the networked seceder model is displayed in
Fig.~\ref{fig:deg}. We note that $P(k)$ has an exponential tail,
notably larger than the Poisson degree distribution~\cite{doromen:rev}
\begin{equation}\label{eq:poi}
P(k)=e^{-\bar{k}}\frac{\bar{k}^k}{k!}
\end{equation}
(where $\bar{k}=2M/N$ is the mean degree) of the initial
random graph, but far from as wide as a power-law. Clearly this falls
into one of the cases mentioned above. We note that as $p$ grows the
degree distribution gets closer to the original network (this was
anticipated in Sect.~\ref{sec:commu}).

\begin{figure}
  \centering{\resizebox*{0.9 \linewidth}{!}{\includegraphics{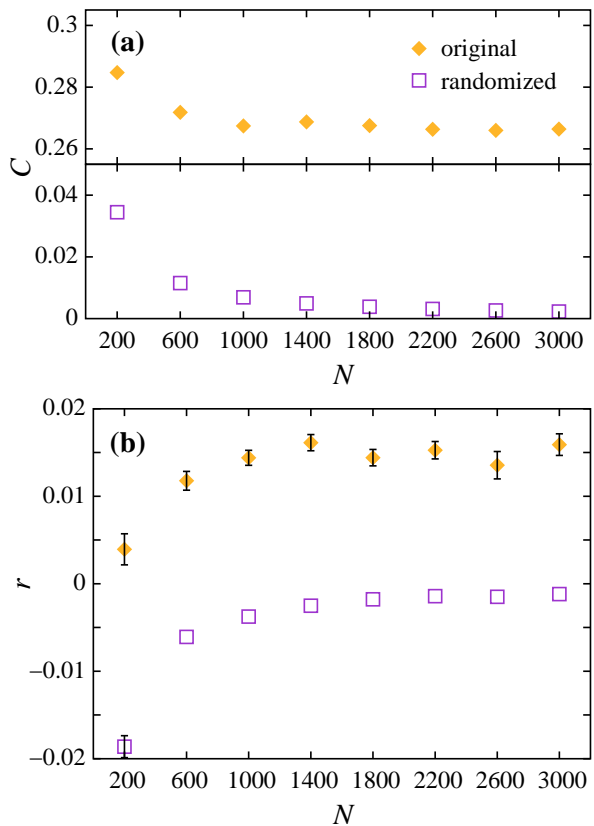}}}
  \caption{Common structural measures. (a) shows the clustering
  coefficient as a function of the number of vertices for the seceder
  model and rewired networks. (b) displays the corresponding plot of
  the assortative mixing coefficient. The network parameters are
  $M=3N$ and $p=0.1$. Error bars are shown if they are larger
  than the symbol size.} 
  \label{fig:cr}
\end{figure}

\subsection{Clustering coefficient}

The clustering coefficient $C$ measures the fraction of connected triples
of vertices that form a triad. This type of statistics has been
popular since Ref.~\cite{wattsstrogatz}. The definition we use is
slightly different from that of Ref.~\cite{wattsstrogatz}:
\begin{equation}
C=\frac{c(3)}{p(3)}~,
\end{equation}
where $c(n)$ denotes ten number of representations of circuits of
length $n$ and $p(n)$ denotes the number of representations of paths
of length $n$. (By `representation' we mean an ordered triple such
that one vertex is adjacent to the vertex before or after. For
example, a triangle has six representations---all permutations of the
three vertices.) This definition is common in sociology (although
sociologists emphasize triad statistics for directed networks)---see
Ref.~\cite{leik:book} for a review---but is also frequent in
physicists' literature since Ref.~\cite{bw:sw}. A plot of $C$ as a
function of $N$ is shown in Fig.~\ref{fig:cr}(a). We see that $C$ for
the seceder model converges to a constant value rather
rapidly. Similarity the $C$ for the rewired networks goes to zero
roughly over the same time scale. The fact that community structure
induces a high clustering is well known and
modeled~\cite{mejn:clumodel}, as is the fact that the clustering
vanishes like $1/N$ in a random graph with Poisson degree
distribution~\cite{mejn:rev}.

\subsection{Assortative mixing coefficient}

The assortative mixing coefficient~\cite{mejn:assmix} is the Pearson
correlation coefficient of the degrees at either side of an edge:
\begin{equation}
  r=\frac{4\langle k_1\, k_2\rangle -
    \langle k_1 + k_2\rangle^2}
  {2\langle k_1^2+k_2^2\rangle - \langle
    k_1+ k_2\rangle^2}
\end{equation}
where subscript $i$ denotes the $i$th argument and average is over
the edge set. $r$ is known to be positive in many social
networks~\cite{mejn:assmix,mejn:mix} (networks of online interaction
does not seem to follow this rule~\cite{pok}). It has been suggested
that this  assortative mixing can be related to community
structure~\cite{mejn:why}. Against this backdrop it is pleasing, but
not surprising, to note that the networked seceder model produces
networks with markedly positive $r$, see Fig.~\ref{fig:cr}(b). The
reference networks with the same degree sequences converges to zero
from negative values, as also observed in Ref.~\cite{pok}. It has
been argued~\cite{maslov:inet,mejn:oricorr} that networks formed by
agents without any preference for the degrees of the neighboring
vertices gets negative $r$ from the restriction that only one edge can
go between one pair of vertices. This is probably the reason for the
negative $r$ values of the rewired networks.

\section{Characteristics of community dynamics}

\begin{figure}
  \centering{\resizebox*{0.8 \linewidth}{!}{\includegraphics{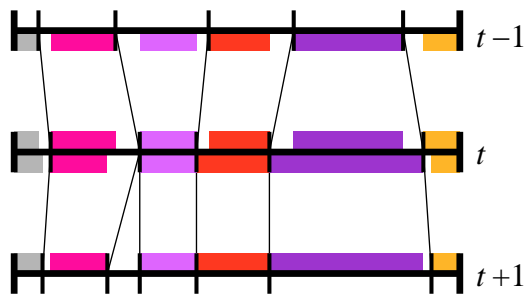}}}
  \caption{Illustration of the identification of clusters at
    consecutive time steps. The vertex set is represented by the
    horizontal line. The vertical tics demarcate cluster
    boundaries. The communities at consecutive time step is matched so
    that the overlap (the horizontal sum of shaded segments) is maximized.} 
  \label{fig:part}
\end{figure}

\begin{figure*}
  \centering{\resizebox*{0.8 \linewidth}{!}{\includegraphics{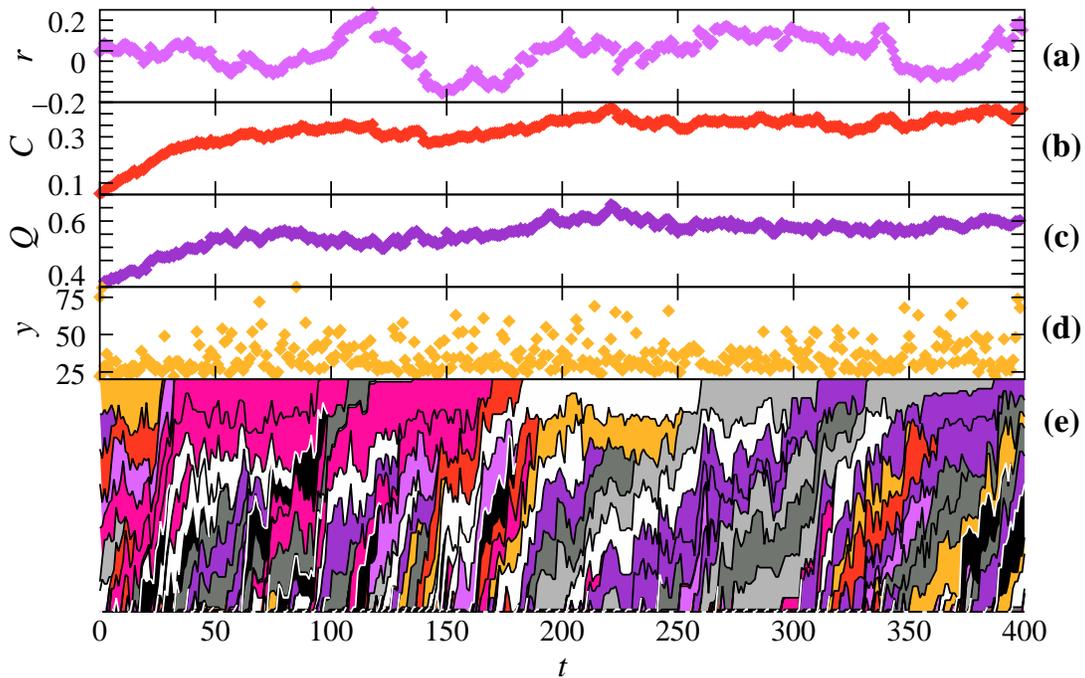}}}
  \caption{Community dynamics for a typical run with the parameter
    values $N=100$, $M=300$ and $p=0$ and 100 iterations of the
    networked seceder model. The different panels shows different
    statistics for one single run of the algorithm. (a) shows the time
    evolution of the assortative mixing  coefficient. (b) shows the
    clustering coefficient $C$. (c) shows the modularity $Q$. (d)
    shows the maximal overlap $y$ between consecutive time steps. (e)
    illustrates the time evolution of the communities. A vertical
    cross section of (e) gives the respective relative sizes of the
    different clusters. The clusters are sorted horizontally according
    to age---the oldest clusters are in the top of the panel.}
  \label{fig:tme}
\end{figure*}

In this section we look at the dynamics of the communities. To do this
we need criteria for if a cluster $B_k^t$ at time $t$ is the same as
cluster $B_{k'}^{t-1}$ at time $t-1$. The idea is to find the best
possible matching of vertices between the partition into clusters of
the two consecutive time steps. To give a mathematical definition, let
$\mathcal{B}_t=\{B_t^1,\cdots,B_t^{b(t)}\}$ be the partition of $G_t$
into clusters by the algorithm described in Sect.~\ref{sec:clust} and
let $b'=\min(b(t),b(t-1))$. Now we define a mapping $f$ from $b'$
elements of $[1,b(t-1)]$ to $b'$ elements of $[1,b(t)]$ such that the
overlap
\begin{equation}\label{eq:clcond}
y'_t=\sum_{k=1}^{b'} |B^k_{t-1} \cap B^{f(k)}_t|
\end{equation}
is maximized ($|\:\cdot\:|$ denotes cardinality). Let $y(t)$ denote this
maximized $y'_t$ value. To calculate this overlap we use the
straightforward method of testing all matchings. In principle this
algorithm runs in exponential time, but since the number of groups is
typically rather low systems of a few hundred vertices numerically
tractable.

The evolution of the group structure, with the group structure
identified as described above, is displayed in Fig.~\ref{fig:tme}. In
Figs.~\ref{fig:tme}(a) and (b) we see the time evolution of the
assortative mixing coefficient $r$ and the clustering coefficient
$C$, whose average size-scaling was studied in
Sect.~\ref{sec:osc}. We note that the assortative mixing 
coefficient fluctuates rather much. Even though it is mostly positive
(remember that the average value is significantly positive) it also
have rather pronounced negative values. This is likely to be a finite
size phenomenon---as the assortative mixing coefficient is
self-averaging~\cite{pok}, larger systems would not fluctuate much and
have stable positive values (as seen in Fig.~\ref{fig:cr}(b)). The
clustering coefficient as displayed in Fig.~\ref{fig:tme}(b) shows a
more stable evolutionary trajectory. Over a time scale roughly
corresponding to $N=100$ updating steps $C$ goes from the value of the
initial Erd\"{o}s-R\'{e}nyi value to the higher clustering coefficient
of the networked seceder model. This is natural since it is also
roughly the time scale for all vertices to be picked and rewired
once. The value of the modularity $Q$, displayed in
Fig.~\ref{fig:tme}(c), is shows a similar behavior as the clustering
coefficient as it increases from the value $\sim 0.4$ of the original
random graph to $\sim 0.6$ of the seceder model. $C$ and $Q$ seem to
be strongly correlated, something that seems very logical in the
context of the seceder model---the clustering coefficient increases
when a high degree vertex is rewired to a specific cluster, a process
that also strengthens the community structure. If this strong $C$-$Q$
correlation is a more ubiquitous property is an interesting problem
for future studies. In Fig.~\ref{fig:tme}(d) we plot the overlap $y$
which fluctuates between $25$ and $75$ with an average well below
$50$. These values are lower than we expected \textit{a priori}, as it
means than identity of more than half the group members changes a
typical time step. Just as the fluctuations in $r$, we expect the
fluctuations in the cluster structure to decrease with system size,
therefore $y/N$ will increase with $N$. In Fig.~\ref{fig:tme}(e) the
time development of different cluster sizes is illustrated. A
horizontal cross section gives the size partitioning of the vertex
set at a given time step. A demarcated area represents a group. Older
groups are above younger groups. An observation from
Fig.~\ref{fig:tme}(e) is that groups typically lives between one and
100 time steps. The life-time scale of groups seems to coincide with
that of the initial relaxation to the seceder equilibrium. We also
note that there seems to be no particular correlation between age and
stability or size, a situation that would have produced skewed life
time or cluster size distributions. At the bottom of the diagram,
hardly visible, there are numerous small, short-lived,
clusters. This is an effect of isolates constantly present in the
system (for this set of parameter values there are typically one or
two at a time step).

The observations in this section were checked for a few other runs and
seems to be representative. Since they do not hint some surprising
phenomena (against the backdrop of the previous sections and the
algorithm itself) we do not conduct any extensive statistical survey
of the dynamical properties.

\section{Summary and conclusions}

We have proposed a model for network formation based on the seceder
model. The model captures how a community structure can emerge from
the desire to be different, both in social and economic systems. The
community structure of our model is analyzed with a recent graph
clustering scheme. This scheme has the advantage that it gives a
measure of the degree of community structure in a network---the
modularity $Q$. We see that the $Q$ is much higher for our model
networks than for random reference networks with the same degree
distributions. Both the number of groups and the average of size of
groups growth as power-laws with sub-linear exponents. Both the
average geodesic distance between vertices of the same and different
clusters grows logarithmically; the difference between these, however,
is much larger for the networked seceder model than for the random
reference networks. The general picture is thus the the networked
seceder model generates well-defined communities just like the agents
of the original seceder model gets clustered in trait space.

The networked seceder model gives networks of high clustering and
positive assortative mixing by degree---properties that are known to
be characteristic of acquaintance networks. The degree distribution
has a peak around the average degree and a exponentially decaying,
also that consistent with real world observations.

The dynamics of the communities were briefly investigated by defining a
mapping between consecutive time steps that maximizes an overlap
function. Using this method we conclude that the speed of the dynamics
is set by the size of the system. We see that the clustering
coefficient and modularity are strongly correlated and that older
groups are not necessarily larger than younger.

To epitomize, the networked seceder model gives an mechanism
of emergent community structure that is different from earlier
proposed mechanisms in network models~\cite{motter:sn,jin:mo,skyrms:mo}.
The mechanism is arguably present in,
at least, social networks~\cite{kamp:at}. We speculate that this model
can be applied to networks of companies that are linked if they are
active in the same market.

\section*{Ackowledgenments}

Thanks are due to Beom Jun Kim, Fredrik Liljeros, Petter Minnhagen and
Mark Newman. The authors were partially supported by the Swedish
Research Council through contract no.\ 2002-4135.


\begin{thebibliography}{10}

\bibitem{aiello}
W.~Aiello, F.~Chung, and L.~Lu, \textit{A random graph model for massive
  graphs}, in Proceedings of the 32nd Annual ACM Symposium on Theory of
  Computing, New York, 2000, Association of Computing Machinery, pp.~171-180.

\bibitem{ba:rev}
R.~Albert and A.-L. Barab\'{a}si, \textit{Statistical mechanics of complex
  networks}, Rev. Mod. Phys \textbf{74} (2002), pp.~47-98.

\bibitem{alb:attack}
R.~Albert, H.~Jeong, and A.-L. Barab\'{a}si, \textit{Attack and error tolerance
  of complex networks}, Nature \textbf{406} (2000), pp.~378-382.

\bibitem{amaral:classes}
L.~A.~N. Amaral, A.~Scala, M.~Barth\'{e}l\'{e}my, and H.~E. Stanley,
  \textit{Classes of small-world networks}, Proc. Natl. Acad. Sci. USA
  \textbf{97} (2000), pp.~11149-11152.

\bibitem{ba:model}
A.-L. Barab\'{a}si and R.~Albert, \textit{Emergence of scaling in random
  networks}, Science \textbf{286} (1999), pp.~509-512.

\bibitem{bara:www}
A.-L. Barab\'{a}si, R.~Albert, H.~Jeong, and G.~Bianconi, \textit{Power-law
  distribution of the world wide web}, Science \textbf{287} (2000), p.\
  2115.

\bibitem{bw:sw}
A.~Barrat and M.~Weigt, \textit{On the properties of small-world network
  models}, Eur. Phys. J. B \textbf{13} (2000), pp.~547-560.

\bibitem{addh}
P.~S. Bearman, J.~Moody, and K.~Stovel, \textit{Chains of affection: The
  structure of adolescent romantic and sexual networks}.
\newblock preprint submitted to Am.\ J.\ Soc.

\bibitem{ber:mormon}
H.~R. Bernard, P.~D. Kilworth, M.~J. Evans, C.~McCarty, and G.~A. Selley,
  \textit{Studying social relations cross-culturally}, Ethnology \textbf{27}
  (1988), pp.~155-179.

\bibitem{bohu:soc}
R.~D. Bock and S.~Z. Husain, \textit{An adaptation of {H}olzinger's
  {B}-coefficients for the analysis of sociometric data}, Sociometry
  \textbf{13} (1950), pp.~146-153.

\bibitem{bollobas_chung_88}
B.~Bollob\'{a}s and F.~R.~K. Chung, \textit{The diameter of a cycle plus a
  random matching}, SIAM J. Discrete Math. \textbf{1} (1988), pp.~328-333.

\bibitem{bonanno:eco}
G.~Bonanno, G.~Caldarelli, F.~Lillo, and R.~N. Mantegna, \textit{Topology of
  correlation based minimal spanning trees in real and model markets}.
\newblock e-print cond-mat/0211546.

\bibitem{harary}
F.~Buckley and F.~Harary, \textit{Distance in graphs}, Addison-Wesley, Redwood
  City, 1989.

\bibitem{sec2}
P.~Dittrich, \textit{The seceder effect in bounded space}, InterJournal
  (2000), art.~no.\ 363.

\bibitem{sec3}
P.~Dittrich and W.~Banzhaf, \textit{Survival of the unfittest? {T}he seceder
  model and its fitness landscape}, in Advances in Artificial Life (Proceedings
  of the 6th European Conference on Artificial Life, Prague, September 10-14,
  2001), J.~Kelemen and P.~Sosik, eds., Springer, Berlin, 2001, pp.~100-109.

\bibitem{sec1}
P.~Dittrich, F.~Liljeros, A.~Soulier, and W.~Banzhaf, \textit{Spontaneous group
  formation in the seceder model}, Phys. Rev. Lett. \textbf{84} (2000),
  pp.~3205-3208.

\bibitem{doromen:rev}
S.~N. Dorogovtsev and J.~F.~F. Mendes, \textit{Evolution of networks}, Adv.
  Phys. \textbf{51} (2002), pp.~1079-1187.

\bibitem{bornholdt:email}
H.~Ebel, L.-I. Mielsch, and S.~Bornholdt, \textit{Scale-free topology of e-mail
  networks}, Phys. Rev. E \textbf{66} (2002), art.~no.\ 035103.

\bibitem{er:on}
P.~Erd\"{o}s and A.~R\'{e}nyi, \textit{On random graphs {I}}, Publ. Math.
  Debrecen \textbf{6} (1959), pp.~290-297.

\bibitem{fararo}
T.~J. Fararo and M.~H. Sunshine, \textit{A study of a biased friendship net},
  Syracuse University Press, Syracuse, NY, 1964.

\bibitem{gir:alg}
M.~Girvan and M.~E.~J. Newman, \textit{Community structure in social and
  biological networks}, Proc. Natl. Acad. Sci. USA \textbf{99} (2002),
  pp.~7821-7826.

\bibitem{jazz}
P.~Gleiser and L.~Danon, \textit{Community structure in jazz}.
\newblock e-print cond-mat/0307434.

\bibitem{grano:weak}
M.~S. Granovetter, \textit{The strength of weak ties}, Am. J. Sociol.
  \textbf{78} (1973), pp.~1360-1380.

\bibitem{gui:sesi}
R.~Guimer\`{a}, L.~Danon, A.~D\'{\i}az-Guilera, F.~Giralt, and A.~Arenas,
  \textit{Self-similar community structure in organisations}.
\newblock e-print cond-mat/0211498.

\bibitem{pok}
P.~Holme, C.~R. Edling, and F.~Liljeros, \textit{Structure and time-evolution
  of the {Internet} community pussokram.com}.
\newblock e-print cond-mat/0210514.

\bibitem{janson}
S.~Janson, T.~{\L}uczac, and A.~Ruci\'{n}ski, \textit{Random Graphs}, Whiley,
  New York, 1999.

\bibitem{jin:mo}
E.~M. Jin, M.~Girvan, and M.~E.~J. Newman, \textit{The structure of growing
  social networks}, Phys. Rev. E \textbf{64} (2001), art.~no.\ 046132.

\bibitem{kamp:at}
C.~Kampmeier and B.~Simon, \textit{Individuality and group formation: The role
  of independence and differentiation}, Journal of Personality and Social
  Psychology \textbf{81} (2001), pp.~448-462.

\bibitem{katz:cug}
L.~Katz and J.~H. Powell, \textit{Probability distributions of random variables
  associated with a structure of the sample space of sociometric
  investigations}, Annals of Mathematical Statistics \textbf{28} (1957),
  pp.~442-448.

\bibitem{leik:book}
R.~K. Leik and B.~F. Meeker, \textit{Mathematical sociology}, Prentice-Hall,
  Englewood Cliffs, NJ, 1975.

\bibitem{liljeros:sex}
F.~Liljeros, C.~R. Edling, L.~A. {Nunes Amaral}, H.~E. Stanley, and
  Y.~{\AA}berg, \textit{The web of human sexual contacts}, Nature \textbf{411}
  (2001), p.\ 907.

\bibitem{maslov:inet}
S.~Maslov, K.~Sneppen, and A.~Zaliznyak, \textit{Pattern detection in complex
  networks: Corelation profile of the {I}nternet}.
\newblock e-print cond-mat/0205379.

\bibitem{mcp:dance}
J.~M. McPherson and J.~R. Ranger-Moore, \textit{Evolution on a dancing
  landscape: Organizations and networks in dynamic blau space}, Social Forces
  \textbf{70} (1991), pp.~19-42.

\bibitem{mcp:bird}
J.~M. McPherson, L.~Smith-Lovin, and J.~Cook, \textit{Birds of a feather:
  Homophily in social networks}, Annual Review of Sociology \textbf{27} (2001),
  pp.~415-444.

\bibitem{motter:sn}
A.~E. Motter, T.~Nishikawa, and Y.-C. Lai, \textit{Large-scale structural
  organization of social networks}, Phys. Rev. E \textbf{68} (2003), art.~no.\
  036105.

\bibitem{mejn:fast}
M.~E.~J. Newman, \textit{Fast algorithm for detecting community structure in
  networks}.
\newblock e-print cond-mat/0309508.

\bibitem{mejn:scicolpnas}
\leavevmode\vrule height 2pt depth -1.6pt width 23pt, \textit{The structure of
  scientific collaboration networks}, Proc. Natl. Acad. Sci. USA \textbf{98}
  (2001), pp.~404-409.

\bibitem{mejn:assmix}
\leavevmode\vrule height 2pt depth -1.6pt width 23pt, \textit{Assortative
  mixing in networks}, Phys. Rev. Lett. \textbf{89} (2002), art.~no.\ 208701.

\bibitem{mejn:mix}
\leavevmode\vrule height 2pt depth -1.6pt width 23pt, \textit{Mixing patterns
  in networks}, Phys. Rev. E \textbf{67} (2003), art.~no.\ 026126.

\bibitem{mejn:clumodel}
\leavevmode\vrule height 2pt depth -1.6pt width 23pt, \textit{Properties of
  highly clustered networks}, Phys. Rev. E \textbf{68} (2003), art.~no.\
  026121.

\bibitem{mejn:rev}
\leavevmode\vrule height 2pt depth -1.6pt width 23pt, \textit{The structure and
  function of complex networks}, SIAM Rev. \textbf{45} (2003), pp.~167-256.

\bibitem{mejn:email}
M.~E.~J. Newman, S.~Forrest, and J.~Balthrop, \textit{Email networks and the
  spread of computer viruses}, Phys. Rev. E \textbf{66} (2002), art.~no.\
  035101.

\bibitem{mejn:commu}
M.~E.~J. Newman and M.~Girvan, \textit{Finding and evaluating community
  structure in networks}.
\newblock e-print cond-mat/0308217.

\bibitem{mejn:why}
M.~E.~J. Newman and J.~Park, \textit{Why social networks are diffrent from
  other types of networks}, Phys. Rev. E \textbf{68} (2003), art.~no.\ 036122.

\bibitem{mejn:oricorr}
J.~Park and M.~E.~J. Newman, \textit{Origin of degree correlations in the
  {I}nternet and other networks}, Phys. Rev. E \textbf{68} (2003), art.~no.\
  026112.

\bibitem{radi:comm}
F.~Radicchi, C.~Castellano, F.~Cecconi, V.~Loreto, and D.~Parisi,
  \textit{Defining and identifying communities in networks}.
\newblock e-print cond-mat/0309488.

\bibitem{roberts:mcmc}
J.~M. {Roberts Jr.}, \textit{Simple methods for simulating sociomatrices with
  given marginal totals}, Soc. Netw. \textbf{22} (2000), pp.~273-283.

\bibitem{alon}
S.~Shen-Orr, R.~Milo, S.~Mangan, and U.~Alon, \textit{Network motifs in the
  transcriptional regulation network of {E}scherichia coli}, Nature Genetics
  \textbf{31} (2002), pp.~64-68.

\bibitem{skyrms:mo}
B.~Skyrms and R.~Freemantle, \textit{A dynamic model of social network
  formation}, Proc. Natl. Acad. Sci. USA \textbf{97} (2000), pp.~9340-9346.

\bibitem{sec4}
A.~Soulier and T.~Halpin-Healy, \textit{The dynamics of multidimensional
  secession: Fixed points and ideological condensation}, Phys. Rev. Lett.
  \textbf{90} (2003), art.~no.\ 258103.

\bibitem{taro}
K.~{Taro Greenfeld}, \textit{Speed tribes: Days and nights with {Japan's} next
  generation}, Perennial, New York, 1995.

\bibitem{watts:small2}
D.~J. Watts, \textit{Networks, dynamics, and the small world phenomenon}, Am.
  J. Sociol. \textbf{105} (1999), pp.~493-592.

\bibitem{wattsstrogatz}
D.~J. Watts and S.~H. Strogatz, \textit{Collective dynamics of {`small-world'}
  networks}, Nature \textbf{393} (1998), pp.~440-442.

\end{thebibliography}
\end{document}